\documentclass[12pt]{article}
\usepackage{amsmath,epsfig,latexsym,amssymb,axodraw,color}  
\def\ltsim{\lower3pt\hbox{$\, \buildrel < \over \sim \, $}}  
\def\gtsim{\lower3pt\hbox{$\, \buildrel > \over \sim \, $}}  
\def\be{\begin{equation}}  
\def\ee{\end{equation}}  
\def\ba{\begin{eqnarray}}  
\def\ea{\end{eqnarray}}  
\def\ga{\mathrel{\raise.3ex\hbox{$>$\kern-.75em\lower1ex\hbox{$\sim$}}}}  
\def\la{\mathrel{\raise.3ex\hbox{$<$\kern-.75em\lower1ex\hbox{$\sim$}}}}

\newcommand{\de}{\partial}  
\newcommand{\no}{\noindent} 
\begin{document}

\baselineskip=16pt   
\begin{titlepage}  
\rightline{OUTP-01-24P}  
\rightline{SNS-PH-01-10} 
\rightline{hep-th/0105255}  
\rightline{May 2001} 
\begin{center}  
  
\vspace{0.5cm}  
  
\large {\bf Radion in Multibrane World }  
  
\vspace*{5mm}  
\normalsize  
  
{\bf Ian I. Kogan$^a$\footnote{i.kogan@physics.ox.ac.uk}, Stavros  
Mouslopoulos$^a$\footnote{s.mouslopoulos@physics.ox.ac.uk},  Antonios  
Papazoglou$^a$\footnote{a.papazoglou@physics.ox.ac.uk}\\ and Luigi Pilo$^b$\footnote{pilo@cibslogin.sns.it}}  
  
\smallskip   
\medskip   
{\it $^a$Theoretical Physics, Department of Physics, Oxford University}\\  
{\it 1 Keble Road, Oxford, OX1 3NP,  UK}  
\smallskip    
  
\medskip   
{\it $^b$Scuola Normale Superiore, Piazza dei Cavalieri 7, 56126 Pisa, Italy  
$\&$ INFN} 
\smallskip

\vskip0.6in \end{center}  
   
\centerline{\large\bf Abstract}  
 
The radion dynamics related to the presence of moving  
branes with both positive or negative tensions is studied in the linearized 
approximation. The radion effective Lagrangian is computed for a compact  
system with three branes and  in particular we examine the decompactification  
limit when one brane is sent to infinity. In the non-compact case we  
calculate the coupling of the gravitational modes  
(graviton, dilaton and radion) to matter on the branes.  
The character of gravity on the two branes for all possible combinations of  
brane tensions is also discussed. It turns out that one can have  a  
normalizable dilaton mode even in the non-compact case.  Finally, we   
speculate on the role of moving branes as a possible source of radion  
emission.  
   
\vspace*{2mm}   

\end{titlepage}  
  
\section{Introduction}  
  
The brane universe scenario has been a quite old idea 
\cite{Akama:1982jy,Rubakov:1983bb,Visser:1985qm,Squires:1986aq} but 
recently it has been subject of renewed interest \cite{Arkani-Hamed:1998rs,Antoniadis:1998ig,Arkani-Hamed:1999nn} with the realization 
that such objects are common in string theory. In particular, there has  
been  a lot of activity  on warped brane  constructions in five spacetime 
dimensions, motivated by heterotic M-theory 
\cite{Horava:1996qa,Witten:1996mz,Horava:1996ma} and its five 
dimensional reduction \cite{Lukas:1999yy,Lukas:1999qs}. In the context  
of these constructions, one can localize gravity on the brane world 
having four dimensional gravity even with an extra dimension of 
infinite extent \cite{Gogberashvili:1999tb,Randall:1999vf} (RS2), or can 
generate an exponential mass hierarchy on a compact two brane model as it 
was done in the Randall - Sundrum (RS1) model \cite{Randall:1999ee},  
providing a novel geometrical resolution of the Planck hierarchy 
problem. For more details on warped 
models see an excellent review of \cite{Rubakov:2001kp}. 
 It was soon realized that similar multibrane   
constructions could provide hints to address other long standing puzzles of  
particle physics as fermion mass hierarchies 
\cite{Dvali:2000ha,Libanov:2001uf,delAguila:2000kb} or neutrino 
phenomenology 
\cite{Dienes:1999sb,Arkani-Hamed:1998vp,Mohapatra:1999zd,Grossman:2000ra,Lukas:2000wn,Lukas:2001rg,Mouslopoulos:2001uc}
(both in warped and unwarped approaches) 
by considering bulk fields with non-trivial profiles in the extra dimension. Graviton loop effects on brane observables have been studied in  
\cite{loops}. Moreover, multibrane constructions can entirely change  
our usual conception about gravity by the possibility of having  
ultra-light massive gravitons contributing substantially to  
intermediate distance gravity, a scenario called multigravity  
\cite{Kogan:2000wc,Mouslopoulos:2000er,Gregory:2000jc,Kogan:2000cv,Kogan:2000xc}. An other interesting class of 
models with massive gravitons was studied in 
\cite{Dvali:2000hr,Dvali:2001xg}.  
Observation of modifications of gravity at ultra-large scale could be a   
striking signal of such a possibility 
\cite{Binetruy:2000xv,Uzan:2000mz}. The multigravity possibility is 
intimately  connected with the multilocalization of gravity in 
multibrane constructions, a property which can be (with appropriate 
mass terms) common in fields of all 
spins as we will review in a forthcoming publication \cite{loc}. 
  
In this paper we are interested in describing the dynamics of the  
scalar gravitational perturbations for a generic system with three flat  
3-branes embedded in $AdS_5$. These excitations describe  
the effect of the fluctuation of the size of the extra dimension and/or  
of the relative positions of the branes. We will distinguish   
these two kinds of modes by calling the former dilaton 
\cite{Charmousis:2000rg} and the  
latter radions \cite{Charmousis:2000rg,Pilo:2000et}.  
An orbifold symmetry $Z_2$ acting on the extra dimensional coordinate 
as $y \to -y$ is also imposed. When the topology of the extra dimension 
is $S^1$, the compact case, the $Z_2$ action has two fixed points $y=0, \, L$ 
and two of the three branes are sitting on fixed points.  As a result of the 
$Z_2$ symmetry the branes in  $y=0, \, L$ are frozen and we are left with just 
one radion field corresponding to the fluctuation of the position of the 
third brane and the dilaton corresponding to the fluctuation of the 
size of the orbifold.

Radion excitations play an important role in the context of   
multigravity models. A generic problematic feature of multigravity models with  
flat branes is that massive gravitons have extra polarization states which do  
not decouple in the massless limit \cite{Dvali:2000rv}, this is known as the  
van Dam - Veltman - Zakharov discontinuity \cite{vanDam:1970vg,Zakharov}.  
However, an equally generic characteristic of these  
models is that they contain moving branes of negative  
tension.  In certain models the radion can help to recover 4D gravity on the  
brane at intermediate distances. Indeed, the role of the radion associated  
with the negative tension brane is  
precisely to cancel the unwanted  massive graviton polarizations and recover  
the correct tensorial structure of the four dimensional graviton propagator  
\cite{Dvali:2000km,Pilo:2000et}, something also seen from the bent 
brane calculations of \cite{Csaki:2000ei,Gregory:2000iu}.  
This happens because the radion in this case is a physical ghost because
it is has a wrong sign kinetic term. This fact of course makes the construction  
problematic because the system is probably quantum mechanically unstable. 
Classically,  the origin of the problem  
is the fact that the weaker  energy condition is violated  in the 
presence of moving negative tension branes \cite{Freedman:1999gp,Witten:2000zk}.   
 
A way out of this difficulty is to abandon the requirement of flatness  
of the branes and consider curved ones. A particular example was  
provided in the $''++''$ model of \cite{Kogan:2001vb} where no negative 
tension brane was needed to  get multigravity. The  decompactification limit 
when one of the two branes is sent to infinity was discussed at the same time at \cite{Karch:2001ct} (see also \cite{Miemiec:2000eq,Schwartz:2001ip} for the KK spectrum analysis) and revealed that gravity was  localized on the brane although a graviton zero mode was absent.  Moreover, due to the fact that the branes where  
$AdS_4$ one could circumvent at least at tree level the van Dam -  
Veltman - Zakharov theorem  \cite{Kogan:2001uy,Porrati:2001cp} and the  
extra polarizations of the massive gravitons where practically  
decoupled. Of course one can ask the question about the resurrection
of these extra polarizations in  quantum loops.  One loop effects in
the massive graviton propagator in $AdS_4$ were discussed  in 
\cite{Dilkes:2001av,Duff:2001zz}. 
Of course, purely four-dimensional theory with massive
graviton is  not  well-defined  and it 
 is certainly true that if the mass term is 
added by hand in purely four-dimensional theory a lot of problems will 
emerge as it was shown in the classical paper of \cite{Boulware:1972my}.  If however  the  underlying theory is  a 
higher dimensional one, the graviton(s)
mass terms appear dynamically and this is a different story. 
All quantum corrections must be calculated in a 
higher-dimensional theory, where a larger number of
graviton degrees of freedom is present naturally (a massless
five-dimensional graviton has the same number of  degrees of freedom
as a massive four-dimensional one). 

  Moreover, the smoothness
of the limit $m \rightarrow  0$ 
  is not only a property of the $AdS_4$ space but holds 
for any background where the characteristic curvature invariants are  non-zero
\cite{Iantalk,DamIan,Oxf}. For physical processes taking place 
 in some region a curved space with a characteristic average curvature, 
the effect of graviton mass is controlled by positive powers of the ratios 
 $m^2/R^2$  where $R^2$ is a characteristic curvature invariant (made
from Riemann and Ricci tensors or scalar curvature).
  A very interesting  argument  supporting  the
conjecture that  there is a smooth limit for phenomenologically 
 observable amplitudes in brane gravity  with ultralight gravitons 
 is based on a very interesting paper   \cite{Vain}\footnote{
Unfortunately this paper was unknown (at least  to  us) 
 when the debate about possibility to have  modification of gravity at 
large scales  started more than a year ago and 
  we became aware about it only  recently  from the preprint \cite{Deffayet:2001pu}.}.
  In that paper it  was shown  that there is a smooth
limit for a   metric  around a spherically symmetric  source with a
mass  $M$   in a theory with
massive graviton  with  mass $m$ for  small (\textit{i.e.} smaller than 
 $m^{-1}(mM/M_P^2)^{1/5}$) distances.
   The discontinuity reveals itself at large distances. The
non-perturbative  solution discussed in \cite{Vain} was found in a limited range of distance from
the center and it is still unclear if it can be
smoothly continued to spatial infinity (this problem was stressed in 
 \cite{Boulware:1972my}). Existence of this smooth
continuation depends on the full  nonlinear structure of the
theory. If one adds a mass term by hand the smooth asymptotic at
infinity may not exit. As far as we know this 
question is still open and the only other reference about which we are 
aware is \cite{Goldhaber:1974wg}. However, it 
 seems  plausible   that 
 in all cases when modification of gravity at large distances comes
from consistent higher-dimensional models, the global  smooth solution 
can exist because in this case there is 
a unique non-linear structure related to the mass term  which is
dictated by the  underlying 
higher-dimensional theory. In  a forthcoming paper \cite{DDGV} an example of a
 5d cosmological solution  will be discussed which contains an explicit 
interpolation between perturbative and non-pertubative regimes: a direct analog
of large and small distances in the Schwarschild case\footnote{IIK is
grateful  to A. Vainshtein for a very interesting discussion on this
subject and for informing us about \cite{DDGV}.}.

 An interesting feature of the above $''++''$ model, which only has a 
dilaton, is that the dilaton survives in the decompactification limit \cite{Karch:2001ct,Miemiec:2000eq,Schwartz:2001ip}
when one of the two branes is sent to infinity (see  
\cite{Papazoglou:2001ed,Chacko:2001em} for detailed analysis for the dilaton). Indeed, in  \cite{Karch:2001ct} where this limit was firstly discussed it was found that there exists a massive scalar mode in the gravity perturbation  
spectrum. Although it seems strange to 
have a dilaton in an infinite extra dimensional model, it is clear 
that this mode is precisely the remnant of the decompactification 
process of the  compact  $''++''$ model. This happens as we will show also to multibrane models with 
flat branes and is related  to the fact that the radion has opposite localization properties compared to  the ones of the graviton.

The above multibrane constructions in order to be physically acceptable   
should incorporate a mechanism which will stabilize the moduli  
(dilaton, radions) when  
they have positive kinetic energy and will give them some phenomenologically  
acceptable mass. This  can be achieved by considering for example a  
bulk scalar field \cite{Goldberger:1999uk,DeWolfe:2000cp,Kanti:2000rd,Csaki:2001zn} with non trivial bulk  
potential (for the effect of the Casimir force between the branes see \cite{Hofmann:2000cj}). When the radion has negative 
kinetic energy is still not clear whether one can speak about 
stabilization of these systems. They are probably unstable 
at the quantum level and no one has attempted to 
estimate their life-time. A general condition that guarantees  
stabilization of the dilaton in the case of maximally symmetric branes  
was derived in \cite{Papazoglou:2001ed} and restricts the sum of the effective  
tensions of the branes and the leftover curvature of the brane.  The  
moduli stabilization has greater importance in the context of brane  
cosmology where it was found that it played a crucial role in deriving   
normal cosmological evolution on the branes  
\cite{Kanti:1999sz,Csaki:2000mp,Kanti:2000nz}.  A  
non-perturbative analysis of the dilaton  two  
brane models can be found in \cite{Binetruy:2001tc}.

In the present paper we will discuss about the dynamics of the   
dilaton and radions in flat brane models. Firstly, we will examine  
them in a three brane orbifold model which is a generalization of the  
$''+-+''$ multigravity model \cite{Kogan:2000wc}. The decompactification limit is reached moving one of the orbifold fixed points to infinity.  
For a certain range of tension of the moving brane, the  
dilaton decouples and we recover the result of \cite{Pilo:2000et} for the  
radion field. However, there is a range of tensions of the moving brane where   
both the dilaton and the radion are dynamical. In the decompactification   
limit we study the coupling the dilaton and the radion    
to matter on the branes for all possible combinations of tensions. Finally,  
we will speculate about the effect of the motion of the brane and propose  
that is source of the radion field.

\section{The general three three-Brane system}   
 
\begin{figure}[t]  
\begin{center}  
\begin{picture}(300,200)(0,50)  
\SetOffset(0,-20)  
  
\SetWidth{2}  
\SetColor{Red}  
\Line(150,80)(150,250)  
  
\SetColor{Blue}  
\Line(240,80)(240,250)  
\Line(60,80)(60,250)  
  
\SetColor{Green}  
\Line(-30,80)(-30,250)  
\Line(330,80)(330,250)  
  
\SetWidth{0.5}  
\SetColor{Black}  
\Line(-30,160)(330,160)

\Text(150,60)[c]{$y=0$}  
\Text(60,60)[c]{$y=-r$}  
\Text(240,60)[c]{$y=r$}  
\Text(-30,60)[c]{$y=-L$}  
\Text(330,60)[c]{$y=L$}  
\Text(195,180)[c]{$\Lambda_1$}  
\Text(285,180)[c]{$\Lambda_2$}  
\Text(105,180)[c]{$\Lambda_1$}  
\Text(15,180)[c]{$\Lambda_2$}

\end{picture}  
\end{center}  
  
\caption{General three 3-brane model on an orbifold.}  
\label{model}  
  
\end{figure}
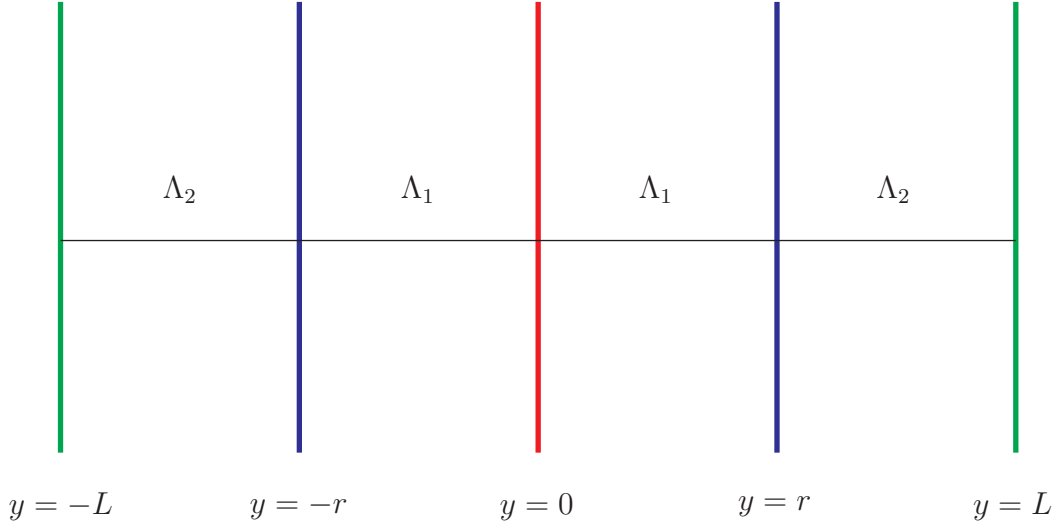

We will consider a three three-brane model on an $S^1/Z_2$ 
orbifold. Two of the branes sit on the orbifold fixed points $y=y_0=0$, 
$y=y_2=L$ respectively. A  third brane is sandwiched in between at 
position $y=y_1=r$ as in Fig.(\ref{model}). In each region between the 
branes the space is $AdS_5$ and  in general the various $AdS_5$ regions  
have  different cosmological constants $\Lambda_1$, $\Lambda_2$. The 
action describing the above system is:  
 
\begin{equation} 
S = \int d^4xdy \sqrt{-G} \left[ 2 M_5^3 \, R \,- \, \Lambda(y)  
\, - \, \sum_i V_i \, \delta(y - y_i)  \,   
\frac{\sqrt{-\hat{g}}}{ \sqrt{-G}} \right] \quad.  
\label{act}  
\end{equation} 
where $M_5$ is the five dimensional Planck mass, $V_i$ the tensions of 
the gravitating branes and $\hat{g}_{\mu\nu}$ the induced metric on the  
branes. The orbifold symmetry $y \to -y $ is imposed.

We seek a background static solution of Einstein equations for the 
following 4D Poincar\'{e} invariant metric ansatz: 
\begin{equation} 
ds^2 = a^2(y) \, \eta_{\mu \nu} dx^\mu dx^\nu \, + \, dy^2  \quad ;  
\label{back}  
\end{equation} 
 
The solution for the warp factor has the usual exponential form:   
\begin{equation} 
a(y) = \begin{cases} e^{-k_1 y } &  ,0 < y < r  \\[0.3cm]  
e^{-k_2 y + r(k_2 - k_1)} & ,r < y < L   
\end{cases} 
\label{sol}  
\end{equation} 
where $k_1$ and $k_2$ are the curvatures of the bulk in the two 
regions and are related to the bulk cosmological constants as:  
\begin{equation}  
k_1^2 = -\frac{\Lambda_1}{24 M_5^3}  \, , \qquad  
k_2^2= -\frac{\Lambda_2}{24 M_5^3} \; .  
\end{equation} 
 
Moreover, the Einstein equations impose the following fine tuning 
between the brane tensions and the bulk cosmological constants: 
 
\begin{equation} 
V_0=24 M_5^3 \, k_1 \, , \qquad V_1=24 M_5^3 \frac{(k_2-k_1)}{ 2} \, , 
\qquad V_2=-24 M_5^3 \, k_2 \; . 
\end{equation}  
 
It is straightforward to recover some special models 
that have been considered in the literature. The RS1 model is obtained 
for $k_1=k_2$ where the intermediate brane is absent (zero tension).  
For $k_1>0$ and $k_2=-k_1$ we get the $''+-+''$ multigravity model 
\cite{Kogan:2000wc,Mouslopoulos:2000er}. For $k_1>0$ and $k_2>k_1$ we 
get the $''++-''$ brane model considered in \cite{Kogan:2000xc}. In 
the decompactification limit where $L \rightarrow \infty$ we get also 
two interesting models: For  $k_1>0$ and  $k_2=0$ we obtain the 
Gregory - Rubakov - Sybiriakov (GRS) model \cite{Gregory:2000jc} and 
for  $k_1>0$ and $k_2=0>k_1$ the non-zero tension version 
\cite{Pilo:2000et} of the model considered in \cite{Lykken:2000nb}.

\section{Effective action}

Our purpose is to study fluctuations of the background (\ref{back}).  
The first important observation is that there exists a generalization of   
Gaussian normal coordinates such that in the perturbed geometry the embedding  
of branes is still  described by $y=0, \, y=r$ and $y=L$, and the metric has the form
(see for instance the  appendix of \cite{Pilo:2000et}):   
\begin{equation}  
ds^2 = g_{\mu \nu}(x,y) \,  dx^\mu dx^\nu  + g_{yy}(x,y) \, dy^2 \quad .  
\end{equation}  
When analyzed from a 4D point view, in each region,  perturbations are of  
of three types.  
\begin{itemize}  
\item  
{\bf Spin two}:  
 
\no 
Tensor-like perturbation $h_{\mu\nu}(x,y)$ corresponding to massive   
(massless) 4D gravitons   
\begin{equation}  
ds^2 = a^2(y) \left[\eta_{\mu \nu} + h_{\mu \nu}(x,y) \right] \, dx^\mu dx^\nu  
+ dy^2 \quad .  
\end{equation}  
\item  
{\bf Spin zero: Dilaton}   
 
\no 
Scalar perturbation $f_1(x)$ corresponding to an overall rescaling of 
distances \cite{Charmousis:2000rg} 
\begin{equation}  
ds^2 = a^2(y) \left[1+ Q(y) f_1(x) \right] \eta_{\mu \nu} \, dx^\mu dx^\nu   
+ \left[1+ q(y) f_1(x) \right] dy^2 \quad .  
\end{equation}   
\item  
{\bf Spin zero: Radion}  
 
\no 
Scalar perturbation $f_2(x)$ corresponding to a fluctuating distance   
of the branes \cite{Pilo:2000et}  
\begin{equation}  
ds^2 = a^2(y) \left[ \left(1+ B(y) f_2(x) \right) \eta_{\mu \nu} +    
2 \epsilon(y) \, \de_\mu \de_\nu f_2(x) \right]  
dx^\mu dx^\nu   
+ \left[1+ 2 A(y) \,  f_2(x) \right] dy^2  .  
\end{equation}   
\end{itemize}   
The function $\epsilon(y)$ is needed in order that the metric satisfies the Israel junction
conditions in the presense of moving branes. As it will be shown later, the values of 
$\de_y \epsilon$ at $y=0, \, r, L$ are gauge invariant. When $k_1 \neq k_2$ a non-trivial $\epsilon$ is required. 

The generic perturbation can be written as   
\begin{equation}  
\begin{split}  
ds^2 =& a^2(y) \left\{ \left[1+ \varphi_1(x,y) \right] \eta_{\mu \nu} + 2   
\epsilon(y) \, \de_\mu \de_\nu f_2(x) +   
h_{\mu \nu}(x,y) \right \} dx^\mu dx^\nu \,   \\  
& +\left[1+  \varphi_2(x,y) \right] dy^2 \quad ;  
\end{split}  
\label{pert}  
\end{equation}  
where   
\begin{equation}  
\begin{gathered}  
\varphi_1(x,y)  =  Q(y) f_1(x) + B(y) f_2(x)\\  
\varphi_2(x,y)  = q(y) f_1(x) + 2A(y) f_2(x)     
\end{gathered} \quad .  
\label{ans}  
\end{equation}  
Given the expression (\ref{sol}) for $a$, Israel junctions conditions at $y=0,  
\, L$, simply require that $A, \, B, \,  \de_y \epsilon, \, Q, \, q$ are  
continuous there \cite{Pilo:2000et}.   
The 4D effective action $S_{eff}$ for the various modes is obtained inserting   
the ansatz (\ref{ans}) in the action (\ref{act}) and integrating out $y$.   
So far the functions $A, \, B, \, \epsilon, \, Q, \,   
q$ haven't been specified, however imposing that $S_{eff}$ contains no   
mixing terms among $h$ and  $f_i$ one determines $Q, \, q$ and $A$ is  
expressed in terms of $B$ (see appendix) which satisfies  
\begin{gather}  
\frac{d}{dy} \left(B a^2 \right) + 2 a^{-1} \, \frac{da}{dy} \, \frac{d}{dy}   
\left(a^4 \de_y \epsilon \right) = 0 \quad ; \label{bdiff} \\  
 \int_{- L}^{L} dy \; a \left(\frac{da}{dy} \right)^{-1} \,   
\frac{dB}{dy} = 0 \quad . \label{nomix}  
\end{gather}  
As a consequence of the no-mixing conditions the linearized Einstein equations  
for (\ref{ans}) will consist in a set independent equations for the graviton  
and the scalars.  
  
The effective Lagrangian reads (see appendix)  
\begin{equation}  
\begin{split}  
S_{eff} &= \int d^4 x  \, {\cal L}_{eff} =  \int d^4 x  \, \left(  
{\cal L}_{Grav} + {\cal L}_{Scal} \right) \\  
{\cal L}_{eff} &=  2 M_5^3 \int_{- L}^{L} dy  \, \Big \{a^2    
{\cal L}_{PF}(h) +   
\frac{a^4}{4} \left[(\de_y h)^2 - \de_y h_{\mu \nu} \, \de_y h^{\mu \nu}   
\right] + {\cal L}_{Scal}\Big \} \quad  ;  
\end{split}  
\label{eff1}  
\end{equation}  
with  
\begin{equation}  
\begin{split}  
{\cal L}_{Scal} &= {\cal K}_1 \, f_1 \Box f_1 +  {\cal K}_2 \, f_2 \Box f_2 \\  
 {\cal K}_1 =  2 M^3_5 \frac{3}{2} c^2\int_{- L}^{L} a^{-2}   
\, dy \, ,& \quad   
{\cal K}_2 = - 2 M^3_5 \, \frac{3}{4} \int_{- L}^{L} a   
\left(\frac{da}{dy} \right)^{-1} \, \frac{d}{dy} \left( B^2 a^2 \right) \, dy  
\; .  
\end{split}  
\label{kt}  
\end{equation}  
In (\ref{eff1}), the spin-2 part,  as expected,  contains the 4D   
Pauli-Fierz Lagrangian ${\cal L}_{PF}(h)$ for the graviton (see Appendix) plus a mass term  
coming from the dimensional reduction. In the scalar part ${\cal L}_{Scal}$  
the mass terms are zero since $f_i$ are moduli fields. Notice that after the  
no-mix conditions are enforced, the effective Lagrangian  
contains the undetermined function $\epsilon$.  
 
The metric ansatz $G_{{}_{MN}}$ in (\ref{ans}) is related to a special  
coordinate choice (generalized Gaussian normal), nevertheless a residual  
gauge (coordinate) invariance is still present. Consider the class of  
infinitesimal coordinate transformations  $X^M \to {X^\prime}^M= X^M +  
\xi(X)^M$ such that  
the transformed metric ${G_{{}_{MN}}}^\prime = G_{{}_{MN}} + \delta  
G_{{}_{MN}}$ retains the original form (\ref{pert}) up to a  
redefinition of the functions $q,Q,A,B, \epsilon$ and the dilaton 
and the radion field. Consistency with the orbifold geometry and the  
requirement 
that the branes in $y=0$, $r$ and $L$ are kept fixed by diffeormorphisms lead to $ 
\xi^5(x,0) = \xi^5(x,r) = \xi^5(x,L) =0 $. From  
\begin{equation} 
\delta G_{{}_{MN}} = - \xi^A \, \de_A  G_{{}_{MN}} - \de_{{}_M} \xi^A \,  
 G_{{}_{AN}} - \de_{{}_N} \xi^A \, G_{{}_{MA}} \;, 
\end{equation} 
one can show that $\xi^M$ has to be of the form 
\begin{equation} 
\begin{gathered} 
\xi^\mu(x,y) = \hat{\xi}^\mu(x)  - W(y) \, \eta^{\mu \nu} \,  
\de_\nu f_2(x) 
\, , \qquad \xi^5(x,y) = a^2 \, W^\prime(y) \, f_2(x)  \\ 
\text{with } ~ W^\prime(0) =  W^\prime(r) =  W^\prime(L) = 0 \quad . 
\end{gathered} 
\end{equation} 
The case $W = 0$ corresponds the familiar pure 4D diffeormorphisms, under which $h_{\mu \nu}$  
transforms as spin two field, $f_i$ as scalars and $q,Q,A,B, \epsilon$ are 
left unchanged. On the contrary the case $\hat{\xi}^\mu = 0, \; W \neq 0 $ is 
relic of 5D diffeormorphisms and one can check that $q,Q,A,B, \epsilon$ are not 
invariant and in particular $\delta \epsilon = W$.   
As a result, the values of $\de_y \epsilon$ in  
$0$, $r$ and $L$ are gauge independent and this renders   
${\cal L}_{eff}$ free from any gauge ambiguity.

\section{Scalars Kinetic Energy}  
  
\subsection{The compact case}  
In this section we will focus on the part of the effective Lagrangian 
involving the scalars and concentrate on the dilaton and radion kinetic 
term coefficients  ${\cal K}_1$ and ${\cal K}_2$. In particular we are  
interested in the cases when the radion becomes a ghost field, 
\textit{i.e.} ${\cal K}_2<0$. Firstly, it is trivial to obtain the dilaton kinetic term ${\cal K}_1$ by integrating (\ref{kt}):  
  
\begin{equation}  
{\cal 
K}_1=3c^2M_5^3\left[\frac{a^{-2}(r)-1}{k_1}+\frac{a^{-2}(L)-a^{-2}(r)}{k_2}\right]  
\label{K1} 
\end{equation}

\begin{figure}[t]  
\begin{center}  
\begin{picture}(200,200)(0,0)  
\LongArrow(0,100)(200,100)   
\LongArrow(100,0)(100,200)  
\SetColor{Blue}  
\Text(110,190)[lb]{$k_2$}  
\Text(195,80)[lb]{$k_1$}  
\DashLine(0,0)(200,200){10}  
\Text(190,185)[lt]{\Blue{$k_1=k_2$}}  
\Text(50,140)[rb]{\color{red}${\cal K}_2 >0$}  
\Text(150,40)[lt]{\color{red}${\cal K}_2 <0$}  
\SetColor{Red}  
\Text(155,80)[lt]{\color{green} GRS}  
\SetColor{Green} 
\Line(100,100)(195,100) 
\LongArrow(150,80)(140,95)  
\SetColor{Black}  
\Vertex(100,100){2} 
\Text(115,140)[lb]{$A$}  
\Text(85,60)[rt]{$A'$} 
\Text(140,110)[lb]{$B$} 
\Text(60,90)[rt]{$B'$} 
\Text(125,80)[lt]{$C$} 
\Text(75,120)[rb]{$C'$} 
\end{picture}  
\end{center}  
\caption{Sign of the radion kinetic term in the $(k_1,k_2)$ plane. In 
the regions $A$, $B'$, $C'$ the moving brane has positive tension 
and the radion positive kinetic energy. In the regions $B$, $C$, $A'$ 
the moving brane has negative tension and the radion  negative kinetic 
energy. We show the GRS line for the non-compact case. The dashed line  
corresponds to $k_1=k_2$, \textit{i.e.} a \textit{tensionless} moving brane.}  
\label{phase}  
\end{figure}
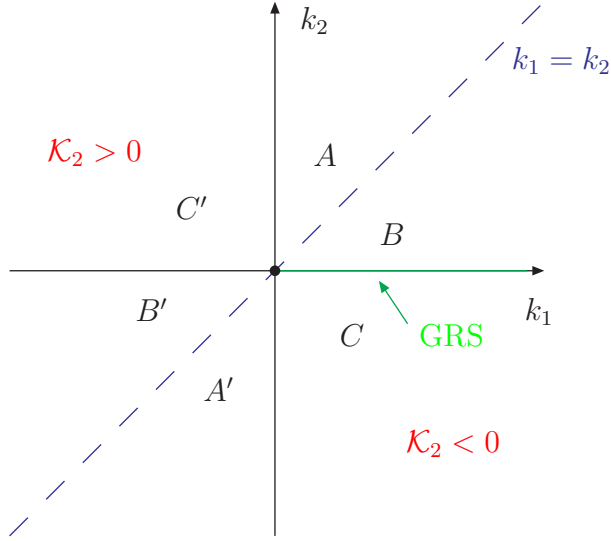

It turns out that for any possible values of $k_1$, $k_2$ and $r$, $L$  
the above quantity is positive definite. The radion kinetic term on 
the other hand is more involved.  
Integrating (\ref{bdiff}) we  get the  radion wavefunction for the  
regions ($y>0$):  
\begin{equation}  
B(y)=   
\begin{cases}    
c_1 \, a^{-2} + 2 k_1 \, \de_y \epsilon \, a^2  &  ,0 < y < r \\[0.3cm]   
c_2 \, a^{-2} + 2 k_2 \, \de_y \epsilon \, a^2  &  ,r < y < L   
\end{cases} \quad ;   
\label{B}   
\end{equation}  
where $c_1$ and $c_2$ are integration constants. The orbifold boundary   
conditions demand that $\de_y \epsilon(0) =\de_y \epsilon(L)= 0$ since  
$\epsilon$ is an even function of $y$. From the non-mixing conditions  
for radion and dilaton (\ref{nomix}) and the  
continuity of $B$ we are able to determine $c_2$ and $\de_y  
\epsilon(r)$ as the following:  
\begin{gather}   
c_2=c_1 \frac{k_2}{k_1}~\frac{a^2(r)-1}{\left(\frac{a(r)}{a(L)}\right)^2-1}\\   
\epsilon^\prime(r)=\frac{c_1k_2}{2k_1(k_2-k_1)a^4(r)}\left[\frac{k_1}{k_2}- 
\frac{a^2(r)-1}{\left(\frac{a(r)}{ a(L)}\right)^2-1}\right] \; .  
\end{gather}   
Therefore, the values of the radion  wavefunction $B$ at the brane positions 
are given by the following expressions:  
\begin{gather}  
B(0)=c_1 \; ;\\  
B(r)=\frac{c_1k_2}{(k_2-k_1)} \frac{1-a^2(L)}{a^2(L)\left[\left( 
\frac{a(r)}{ a(L)}\right)^2-1\right]} \; ;\\  
B(L)=\frac{c_1k_2}{k_1}\frac{a^2(r)-1}{a^2(L)\left[\left(\frac{a(r)}{  
a(L)}\right)^2-1\right]} \; \; . 
\end{gather}  
Thus we can carry out the integral in   
(\ref{kt}) to find  the radion kinetic term coefficient:  
\begin{align}   
{\cal K}_2&= 3 M_5^3  \left[\left(\frac{1}{  k_1} - \frac{1}{ k_2} \right)  
 B^2(r) \, a^2(r) + \frac{1}{   
k_2} \, B^2(L) \, a^2(L) - \frac{1}{ k_1} \, B^2(0)\right] \nonumber \\  
&= \frac{3 M_5^3c_1^2}{ k_1}  \left\{ \frac{k_2}{  
(k_2-k_1)}\frac{a^2(r)(a^2(L)-1)^2}{a^4(L)\left[\left(\frac{a(r)}{ 
a(L)}\right)^2-1\right]^2}+\frac{k_2}{  
k_1}\frac{(a^2(r)-1)^2}{a^2(L)\left[\left(\frac{a(r)}{  
a(L)}\right)^2-1\right]^2}-1 \right\} \, .  
\label{K2} 
\end{align}  
The above quantity is not positive definite. In particular, it turns out 
that it is positive whenever the intermediate brane has positive 
tension and  negative whenever the intermediate brane has negative 
tension. This result is graphically represented in Fig.(\ref{phase}) 
where the $(k_1,k_2)$ plane is divided in two regions.

\subsection{The non-compact limit} 
It is instructive to discuss the decompactification limit in which the third brane is  sent  to infinity, \textit{i.e.}  
$L \to + \infty$. To examine this limit we distinguish two cases:  
\vskip 0.5truecm  
\centerline{\bf The case $\mathbf{k_2 > 0}$}  
\vskip 0.5truecm  
\noindent  
In this case we have $\displaystyle{a(\infty) \propto 
\lim_{L\rightarrow \infty} e^{-k_2L} =0}$. The dilaton 
kinetic term is trivial since from (\ref{K1}) we obtain ${\cal K}_1 \to \infty$. In other words 
the dilaton decouples  from the 4D effective theory and the condition 
of absence of kinetic mixing between the scalars (\ref{nomix}) plays 
no role. The radion kinetic term coefficient can be read off from (\ref{K2}):   
\begin{equation}  
{\cal K}_2= \frac{3 M_5^3  c_1^2}{k_1} \left[e^{2k_1r} \frac{k_2}  
{(k_2-k_1)}-1  
\right]  
\end{equation}  
This result agrees with the computation of \cite{Pilo:2000et}. It is 
easy to see that the radion has positive kinetic energy when the moving 
brane has positive tension and negative kinetic energy when the 
tension is negative. Indeed, for $0 < k_1 < k_2$, or for $k_2 > 0$ and 
$k_1 < 0$ we have a 
positive brane and positive kinetic energy. On the other hand for $0 < 
k_2 < k_1$  we have a negative tension brane 
and negative kinetic energy. In the limit $k_2 \rightarrow 0$ we get the GRS  
model with  negative kinetic energy as in \cite{Pilo:2000et}. 
\vskip 0.5truecm  
\centerline{\bf The case $\mathbf{k_2 < 0}$}  
\vskip 0.5truecm  
\noindent  
In this case $\displaystyle{a(\infty) \propto 
\lim_{L\rightarrow \infty} e^{-k_2L} \rightarrow \infty}$. This time 
the dilaton plays in the game since ${\cal K}_1$ is finite as seen 
from (\ref{K1}) and has the value: 
\begin{equation}  
{\cal K}_1 = \frac{3 M_5^3 \, c^2 }{2 k_1 |k_2|}  \,   
\left[ e^{2k_1r} (|k_2|+k_1)  - |k_2| \right]  
 \quad   
\end{equation}  
which is manifestly positive definite. The presence of dilaton mode  
is a bit surprising since it describes the fluctuations of the overall size of  
the system which is infinite. Something similar happens in the  
model of \cite{Karch:2001ct} that has a dilaton mode although it is 
non-compact. The dilaton is a remnant of the decompactification 
process of the $''++''$ model \cite{Kogan:2001vb}  and enters in the game 
because the inverse of the warp factor is normalizable. 
 
The radion kinetic term 
coefficient can be obtained from  taking the limit $L \to  
+ \infty$ in (\ref{K2}). We get 
\begin{equation}  
{\cal K}_2 = \frac{3 M_5^3 \,  c^2_1}{k_1} \left[e^{-2 k_1 r} \, \frac{k_2}  
{(k_2-k_1)} - 1 \right]  
\end{equation}  
The same considerations for the compact case applies here. When $k_1 < k_2 < 0$ we have a 
positive tension brane and a positive kinetic energy. On the other hand when  
$k_2 < k_1 < 0$, or for $k_2 < 0$ and $k_1 > 0$ we have a negative tension  
brane and a negative kinetic energy. In the GRS limit $k_2 \rightarrow 0$,  
the radion has negative kinetic energy.

\section{Gravity on the branes}   
  
In this section we will study how the moduli and graviton(s) couple to   
matter confined on the branes. For simplicity we will study the 
non-compact case. For this purpose, we consider a matter Lagrangian 
${\cal L}_m(\Phi_i,\hat{g})$, where we denote generically with  
$\Phi_i$ matter fields living on the branes;  $\hat{g}$ is 
the induced metric on the branes. Our starting point is the following action  
\begin{equation}  
{\cal S}_{5D}=\int d^4xdy \, \sqrt{G} \, [2M_5^3R-\Lambda(y)]-\sum_i\int d^4x\sqrt{\hat{g}}V_i +\int d^4xdy\sqrt{\hat{g}}{\cal L}_m(\Phi_i,\hat{g})  
\end{equation}   
We have already calculated the effective action for the gravity sector   
for the perturbation (\ref{pert}). It is useful to decompose the perturbation  
$h(x,z)$ in terms of a complete set of  eigenfunctions $\Psi^{(n)}(z)$ of the  
graviton kinetic operator (a suitable 4D gauge fixing like de Donder is  
understood)  
\begin{equation} 
h(x,z)=\sum_{n} \Psi^{(n)}(z) \, h_{\mu\nu}^{(n)}(x) \, + \,  
\int dm \, \Psi(y,m) \, h^{(m)}_{\mu\nu}(x) \;    
\end{equation} 
having taken into account both the discrete and the continuum part of 
the spectrum. 
The effective 4D Lagrangian has the following form  
\begin{equation}  
\begin{split} 
{\cal L}_{4D}&= {\cal L}_m(\Phi_i^c,\eta)  + 2M_5^3 \sum_n^{~~~~*} {\cal  
L}_{PF}(h^{(n)}(x)) +{\cal K}_1 \, f_1 \Box f_1 +  {\cal K}_2 \, f_2 \Box  
f_2\\ 
&- \sum_n^{~~~~*}\frac{\Psi^{(n)}(y_{br})}{ 2}  h_{\mu\nu}^{(n)}(x) 
T^{\mu\nu}- \frac{Q(y_{br})}{ 2} f_1 T - \frac{B(y_{br})}{ 2}f_2 T \; ;   
\end{split} 
\end{equation} 
The matter fields have been rescaled $\Phi_i \rightarrow \Phi_i^c$, to make  
them canonically normalized and the energy momentum tensor $T_{\mu\nu}$ is  
defined with respect to the rescaled fields $\Phi_i^c$. By construction the  
induced background metric on the branes is the flat 4D Minkowski metric.  
The asterix denotes that the sum has to be converted into an  
integral for the continuum part of the spectrum. Finally, defining the   
canonical normalized fields
\begin{equation} 
\bar{h}_{\mu\nu}^{(n)}(x)=\sqrt{2M_5^3}h_{\mu\nu}^{(n)}(x) \, , \quad   
\bar{f}_i=\sqrt{2|{\cal K}_i|}f_i \; , 
\end{equation}  
the Lagrangian  reads  
\begin{equation}  
\begin{split} 
{\cal L}_{4D}&= {\cal L}_m(\Phi_i^c,\eta)  + \sum_n^{~~~~*} {\cal  
L}_{PF}(\bar{h}^{(n)}(x)) +\frac{1}{2} \, \bar{f}_1 \Box \bar{f}_1 +  
\frac{1}{ 2} \, \text{sgn}({\cal K}_2) \, \bar{f}_2 \Box \bar{f}_2 \\ 
&-\sum_n^{~~~~*} 
\frac{\Psi^{(n)}(y_{br})}{ 2\sqrt{2M_5^3}}  \, \bar{h}_{\mu\nu}^{(n)}(x) 
\, T^{\mu\nu}  - \frac{Q(y_{br})}{ 2\sqrt{2{\cal K}_1}} \, \bar{f}_1 \, T 
- \frac{B(y_{br})}{  2\sqrt{2|{\cal K}_2|}} \, \bar{f}_2 \, T \quad . 
\end{split} 
\end{equation}   
Thus the dilaton, the radion and the graviton (whenever it is 
normalizable) have the following couplings respectively: 
 
\begin{equation} 
{\cal C}_D={Q(y_{br}) \over 2\sqrt{2{\cal K}_1}} \, , \qquad {\cal C}_R={B(y_{br}) \over 2\sqrt{2|{\cal K}_2|}} \, , 
\qquad {\cal C}_G={\Psi^{(0)}(y_{br}) \over 2\sqrt{2M_5^3}} \; . 
\end{equation}  
For convenience we will define ${\cal C}_N \equiv {\sqrt{|k_1|} \over 
2 \sqrt{2 M_5^2}}$ and write the above quantities as: 
 
\begin{equation} 
{\cal C}_D^{(i)}={\cal C}_N{g_D^{(i)} \over \sqrt{3}} \, , \qquad {\cal C}_R^{(i)}={\cal C}_N{g_R^{(i)} \over \sqrt{3}} \, , 
\qquad {\cal C}_G={\cal C}_N {\Psi^{(0)}(y_{br}) \over \sqrt{|k_1|}} \; . 
\end{equation}  
where the dimensionless radion couplings are: 
 
\begin{eqnarray} 
g_D^{(0)}=\frac{\theta(-k_2)}{\sqrt{\left|e^{2k_1r}{k_1+|k_2| \over 
|k_2|}-1\right|}} \, &,& \qquad g_D^{(r)}=e^{2k_1r}g_D^{(0)}\\ 
g_R^{(0)}=\frac{1}{\sqrt{\left|e^{2k_1r{\rm sgn}(k_2)}{k_2 \over 
k_2-k_1}-1\right|}} \, &,& \qquad g_R^{(r)}=e^{2k_1}{k_2 \over k_2 
-k_1}g_R^{(0)} 
\label{cc} 
\end{eqnarray} 
 
We should note at this point that there are certain cases as we will 
see in the following that multigravity is realized and four 
dimensional gravity in 
intermediate distances is due to more than one mode. In that case we 
will denote by ${\cal C}_G$ the coupling of the ``effective zero 
mode'' even though there might not be a genuine zero mode at all.

\begin{figure}[t]  
\begin{center}  
\begin{picture}(200,150)(0,0)

\Text(-70,130)[lb]{$\sigma(y)$} 
\Text(90,30)[rb]{$y$} 
\LongArrow(-80,0)(-80,140)  
\LongArrow(-100,20)(90,20) 
\Vertex(-80,20){2} 
\Line(-80,20)(-30,70) 
\Text(-75,60)[lb]{$k_1>0$} 
\SetColor{Green} 
\DashLine(-30,70)(90,70){5} 
\Text(70,75)[lb]{\color{green}$k_2=0$} 
\Text(70,65)[lt]{\color{green}GRS} 
\SetColor{Black} 
\DashLine(-30,70)(30,130){5} 
\Text(35,130)[lb]{$k_2=k_1$} 
\SetColor{Red} 
\Line(-30,70)(0,120) 
\Text(5,125)[lb]{\color{red}$A$}  
\SetColor{Blue} 
\Line(-30,70)(50,100) 
\Text(55,100)[lb]{\color{blue}$B$}  
\SetColor{Purple} 
\Line(-30,70)(50,40) 
\Text(55,40)[lt]{\Purple{$C$}} 
 
 
\Text(130,130)[lb]{$\sigma(y)$} 
\Text(290,110)[rt]{$y$} 
\SetColor{Black} 
\LongArrow(120,0)(120,140)  
\LongArrow(100,120)(290,120) 
\Vertex(120,120){2} 
\Line(120,120)(170,70) 
\Text(125,80)[lt]{$k_1<0$} 
\DashLine(170,70)(290,70){5} 
\Text(270,75)[lb]{$k_2=0$} 
\DashLine(170,70)(230,10){5} 
\Text(235,10)[lt]{$k_2=k_1$} 
\SetColor{Red} 
\Line(170,70)(200,20) 
\Text(205,15)[lt]{\color{red}$A'$} 
\SetColor{Blue} 
\Line(170,70)(250,40) 
\Text(255,40)[lt]{\color{blue}$B'$}  
\SetColor{Purple} 
\Line(170,70)(250,100) 
\Text(255,100)[lb]{\Purple{$C'$}} 
 
\end{picture}  
\end{center}  
\caption{The function $\sigma(y)=-\log [a(y)]$ for all possible 
combinations of $k_1$, $k_2$. The regions are named in accordance 
with the Fig.(\ref{phase}) phase diagram.}  
\label{warp}  
\end{figure}
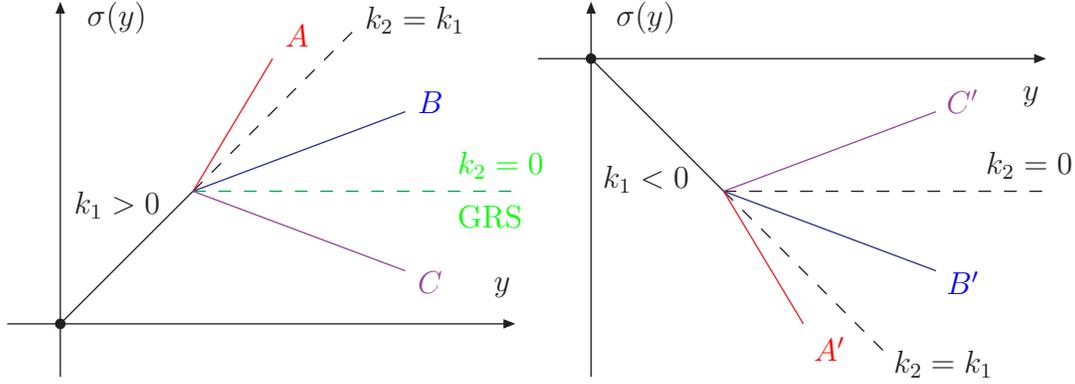

We will now discuss how these couplings behave is the six distinct 
combinations of $k_1$ and $k_2$ shown in Fig.(\ref{warp}).

\vskip 0.5truecm  
\centerline{\bf Region $A$}  
\vskip 0.5truecm  
\noindent  
 
In this case, we have the non-compact system of two positive tension branes discussed in \cite{Lykken:2000nb,Pilo:2000et,Kogan:2000xc}. The volume of the extra dimension is 
finite and therefore we have a normalizable graviton zero mode. The KK 
tower will be continuum and its coupling to matter on either 
branes will be exponentially suppressed for reasonably large $r$ as 
can be easily seen by the analysis of \cite{Kogan:2000xc}. The graviton 
wavefunction in the conformal gauge is constant, \textit{i.e.} 
$\Psi^{(0)}(y)=A$. Thus the universal coupling of the zero 
mode will be ${\cal C}_G={\cal C}_N$.  
 
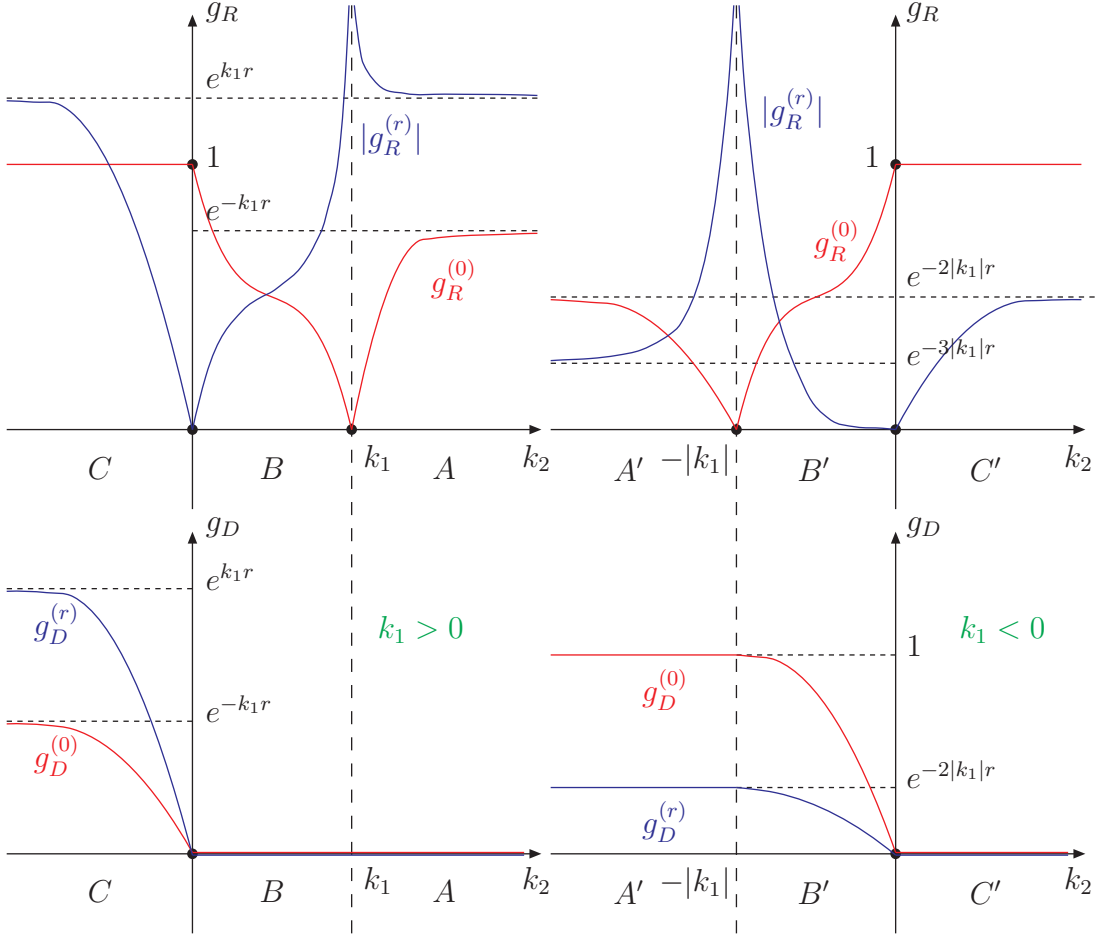
\begin{figure}[t]  
\begin{center}  
\begin{picture}(200,350)(0,0)

\DashLine(95,215)(225,215){2} 
\DashLine(95,240)(300,240){2} 
\DashLine(-40,265)(90,265){2} 
\DashLine(-110,315)(90,315){2} 
 
\DashLine(165,55)(225,55){2} 
\DashLine(-110,80)(-40,80){2} 
\DashLine(165,105)(225,105){2} 
\DashLine(-110,130)(-40,130){2} 
 
\DashLine(165,0)(165,350){5} 
\DashLine(20,0)(20,350){5}

\LongArrow(-40,160)(-40,345)  
\LongArrow(-110,190)(90,190) 
\Vertex(-40,190){2} 
\Vertex(-40,290){2} 
\Vertex(20,190){2} 
\SetColor{Red} 
\Line(-110,290)(-40,290) 
\Curve{(-40,290)(-39,286)(19,194)(20,190)} 
\Curve{(20,190)(45,261)(50,262)(60,263)(70,263.4)(80,263.7)(85,263.9)(90,264)} 
\SetColor{Blue} 
\Curve{(-110,314)(-100,313.5)(-90,312.5)(-40,190)} 
\Curve{(-40,190)(-35,210)(-21,235)(-14,240)(-10,242)(-6,245)(.7,251)(6.2,260)(10.2,270)(18,330)(19,350)} 
\Curve{(21,350)(23.2,333)(24,329)(24.4,328)(26,325)(30,320)(40,316.8)(55,316.5)(70,316.4)(80,316.3)(90,316)} 
\SetColor{Black} 
\Text(-35,345)[lb]{$g_R$} 
\Text(90,185)[tb]{$k_2$} 
\Text(30,185)[tb]{$k_1$} 
\Text(-35,320)[lb]{$e^{k_1r}$} 
\Text(-35,290)[lb]{$1$} 
\Text(-35,270)[lb]{$e^{-k_1r}$} 
\Text(-75,180)[tb]{$C$} 
\Text(-10,180)[tb]{$B$} 
\Text(55,180)[tb]{$A$} 
\Text(35,310)[tb]{$\Blue{|g_R^{(r)}|}$} 
\Text(50,255)[lt]{$\Red{g_R^{(0)}}$}

\LongArrow(-40,0)(-40,150)  
\LongArrow(-110,30)(90,30) 
\Vertex(-40,30){2} 
\SetColor{Red} 
\Line(-40,30.5)(85,30.5) 
\Curve{(-110,79)(-95,78.5)(-85,77)(-40,30.5)} 
\SetColor{Blue} 
\Line(-40,29.5)(85,29.5) 
\Curve{(-110,129)(-95,128.5)(-85,127)(-40,29.5)} 
\SetColor{Black} 
\Text(-35,150)[lb]{$g_D$} 
\Text(90,25)[tb]{$k_2$} 
\Text(30,25)[tb]{$k_1$} 
\Text(-35,130)[lb]{$e^{k_1r}$} 
\Text(-35,80)[lb]{$e^{-k_1r}$} 
\Text(-75,20)[tb]{$C$} 
\Text(-10,20)[tb]{$B$} 
\Text(55,20)[tb]{$A$} 
\Text(-100,125)[lt]{$\Blue{g_D^{(r)}}$} 
\Text(-100,75)[lt]{$\Red{g_D^{(0)}}$} 
\Text(30,120)[lt]{$\Green{k_1>0}$}

\LongArrow(225,160)(225,345)  
\LongArrow(95,190)(295,190) 
\Vertex(225,190){2} 
\Vertex(225,290){2} 
\Vertex(165,190){2} 
\SetColor{Red} 
\Line(225,290)(295,290) 
\Curve{(165,190)(166,194)(224,286)(225,290)} 
\Curve{(95,239)(110,238)(120,237)(165,190)} 
\SetColor{Blue} 
\Curve{(95,216)(100,216.3)(105,216.5)(110,216.8)(120,218)(129,220)(137,224)(140,226)(144,230)(160,300)(162,320)(164,350)} 
\Curve{(166,350)(168,320)(170,300)(191,205)(200,194)(210,191)(220,190.5)(225,190)} 
\Curve{(225,190)(270,238)(280,238.7)(295,239)} 
\SetColor{Black} 
\Text(230,345)[lb]{$g_R$} 
\Text(295,185)[tb]{$k_2$} 
\Text(150,185)[tb]{$-|k_1|$} 
\Text(220,290)[rb]{$1$} 
\Text(230,245)[lb]{$e^{-2|k_1|r}$} 
\Text(230,215)[lb]{$e^{-3|k_1|r}$} 
\Text(125,180)[tb]{$A'$} 
\Text(195,180)[tb]{$B'$} 
\Text(260,180)[tb]{$C'$} 
\Text(175,320)[lt]{$\Blue{|g_R^{(r)}|}$} 
\Text(195,270)[lt]{$\Red{g_R^{(0)}}$}

\LongArrow(225,0)(225,150)  
\LongArrow(95,30)(295,30) 
\Vertex(225,30){2} 
\SetColor{Red} 
\Line(225,30.5)(290,30.5) 
\Line(95,105)(165,105) 
\Curve{(165,105)(170,104.5)(180,103)(225,30.5)} 
\SetColor{Blue} 
\Line(225,29.5)(290,29.5) 
\Line(95,55)(165,55) 
\Curve{(165,55)(170,54.5)(180,53)(225,29.5)} 
\SetColor{Black} 
\Text(230,150)[lb]{$g_D$} 
\Text(295,25)[tb]{$k_2$} 
\Text(150,25)[tb]{$-|k_1|$} 
\Text(230,105)[lb]{$1$} 
\Text(230,55)[lb]{$e^{-2|k_1|r}$} 
\Text(125,20)[tb]{$A'$} 
\Text(195,20)[tb]{$B'$} 
\Text(260,20)[tb]{$C'$} 
\Text(130,100)[lt]{$\Red{g_D^{(0)}}$} 
\Text(130,50)[lt]{$\Blue{g_D^{(r)}}$} 
\Text(250,120)[lt]{$\Green{k_1<0}$}

\end{picture}  
\end{center}  
\caption{The dimensionless couplings of the radion (upper) and the 
dilaton (lower) to matter on the $y=0$ or the $y=r$ branes. The left 
diagrams correspond to $k_1>0$ while the right ones for $k_1<0$. The 
regions $A,~B, \dots$ are in accordance with the ones of the phase 
diagram in   Fig.(\ref{phase}). The diagrams are not in scale.}  
\label{couplings}  
\end{figure}

The dilaton in this region will simply be absent, or equivalently it 
will have zero coupling ${\cal C}_D=0$ to the branes because ${\cal K}_1 \to 
\infty$. The radion in this region has positive kinetic term and from 
(\ref{cc}) we find that its coupling to matter on the $y=0$ 
and $y=r$ branes is bounded as following (see Fig. \ref{couplings}): 
\begin{equation} 
0 \leq g_R^{(0)} \lesssim e^{-k_1r} \, , \qquad g_R^{(r)} \gtrsim e^{k_1r} 
\end{equation} 
 
The radion on the central brane is always weakly coupled and decouples 
in the tensionless moving brane limit. On the other hand the radion is  
always strongly coupled on the moving brane and diverges in the 
tensionless limit where $k_2 \to k_1$.

\vskip 0.5truecm  
\centerline{\bf Region $B$}  
\vskip 0.5truecm  
\noindent

The volume of the extra dimension is still finite and therefore we have a 
normalizable graviton zero mode.  
When $k_1 = k_2$ the brane is tensionless and  as soon as $k_2 < k_1$ 
the KK continuum starts to develop  a resonance. This resonance is initially rather 
broad as $k_2 \to k_1$ and its width decreases until it coincides with 
the GRS width $\Gamma \sim k_1 \exp(-3 k_1 r)$ 
as $k_2 \to 0$. In this region four dimensional 
gravity at intermediate distances is the net effect of the massless 
graviton and the lower part of the KK continuum that contributes more 
and more as $k_2 \to 0$. 
 
The dilaton is again absent, ${\cal C}_D=0$, since still we have 
${\cal K}_1 \to \infty$. The radion in this region is a ghost field, to  
compensate for the presence of the extra polarization states of the 
contributing massive gravitons,  
and its coupling to the  branes is bounded as (see Fig. \ref{couplings}): 
\begin{equation} 
0 \leq g_R^{(0)} \leq 1 \, , \qquad g_R^{(r)} \leq 0 
\end{equation} 
 
The radion on the central brane interpolates between the decoupling 
limit of the tensionless moving brane to the GRS limit where it 
couples with strength equal to the one of the graviton. On the other 
hand the radion coupling on the moving brane interpolates between the 
infinitely strong limit of the tensionless (strictly speaking there is no brane 
in $y=r$ and we are left with RS2) brane to the GRS limit where  
it decouples.

\vskip 0.5truecm  
\centerline{\bf Region $C$}  
\vskip 0.5truecm  
\noindent  
 
In this region we have a infinite analogue of the $''+-+''$ model with  
no normalizable zero mode present. The  
graviton spectrum for $ k_2 \to 0$ is approximately equally 
spaced  and there is a resonance which coincides with 
the GRS width and gets more and more broad as $k_2$ gets more 
negative. Soon enough a special light state is 
singled out as in the $''+-+''$ model, a behaviour that persists for 
all values of $k_2$ in this region. From the above behaviour we 
deduce that  four dimensional gravity at intermediate distances is 
generated by the lower part of the discrete spectrum as  $k_2 \to 0$, 
whereas only by the special state for all other values of $k_2$. 
 
The dilaton in this region is present and its coupling to the branes 
is bounded as following (see Fig. \ref{couplings}): 
\begin{equation} 
0 \leq g_D^{(0)} \lesssim e^{-k_1r}  \, , \qquad 0 
\leq g_D^{(r)}  \lesssim e^{k_1r} 
\end{equation} 
 
It is always weakly coupled to matter on the central brane and becomes 
strongly coupled but saturated on the moving brane. 
 
The radion in this region is again a ghost field to cancel the 
unwanted extra massive graviton polarization states and its coupling to 
the branes is bounded as (see Fig. \ref{couplings}): 
\begin{equation} 
g_R^{(0)} \approx 1 \, , \qquad 0 \leq g_R^{(r)} \lesssim e^{k_1r} 
\end{equation} 
 
On the central brane it couples always with strength equal to the one 
of the ``effective graviton'' and on the moving brane it interpolates 
between the decoupling limit of the GRS case to a strongly coupled 
region with saturated coupling as the tension of the second brane gets  
infinite.

\vskip 0.5truecm  
\centerline{\bf Region $A'$}  
\vskip 0.5truecm  
\noindent

In this region we have no normalizable zero mode and the two negative 
tension brane system resembles an inverted version of the RS2 
model. The KK spectrum  is discrete and all excitations lie above the 
characteristic curvature scale $k_1$. Thus the low energy effective theory does not have four 
dimensional gravity at all. 
 
The dilaton in this system is present and the coupling to the branes 
is approximately constant (see Fig. \ref{couplings}): 
\begin{equation} 
g_D^{(0)} \approx 1  \, , \qquad g_D^{(r)}  \approx 1 
\end{equation} 
 
The radion field is a ghost and its coupling to matter on the branes 
is bounded as (see Fig. \ref{couplings}): 
\begin{equation} 
g_R^{(0)} \gtrsim e^{-3|k_1|r} \, , \qquad 0 \leq g_R^{(r)} \lesssim e^{-2|k_1|r} 
\end{equation} 
 
The radion on the central negative tension brane is always weakly 
coupled and vanishes in the limit of $k_2 \to k_1=-|k_1|$ of the 
tensionless moving brane. On the other hand the coupling on the moving brane 
is bounded from below in the limit of infinite negative tension brane 
and diverges at the limit of the tensionless brane.

\vskip 0.5truecm  
\centerline{\bf Region $B'$}  
\vskip 0.5truecm  
\noindent  
  
In this region there is again no normalizable zero mode and the system  
of the negative and positive tension branes still resembles the 
inverted RS2 model. The KK spectrum is almost identical with the 
one of the previous region except for the limit $k_2 \to 0$ when the 
spectrum drops below the curvature scale $k_1$ and a continuum develops. 
  
The dilaton coupling to the branes decreases from the constant value 
of the previous region, to zero as $k_2 \to 0$ (see Fig. \ref{couplings}): 
\begin{equation} 
0 \leq g_D^{(0)} \lesssim 1  \, , \qquad 0 
\leq g_D^{(r)}  \lesssim e^{-2|k_1|r} 
\end{equation} 
 
The radion has positive kinetic energy and its coupling to the branes 
is bounded as following (see Fig. \ref{couplings}): 
\begin{equation} 
0 \leq g_R^{(0)} \leq 1 \, , \qquad g_R^{(r)} \leq  0 
\end{equation} 
 
It is weakly coupled in the central brane and the coupling interpolates between 
zero from the previous region to one in the inverted GRS case where $k_2 \to 0$. On the other hand  
it is divergent as $k_2 \to k_1=-|k_1|$ and vanishes as $k_2 \to 0$. 
 
\vskip 0.5truecm  
\centerline{\bf Region $C'$}  
\vskip 0.5truecm  
\noindent 
 
In this region we again have a system of a negative and a positive 
tension brane, but gravity  can be localized on the moving positive 
tension brane. There is a normalizable zero mode that mediates four 
dimensional gravity and a KK continuum with suppressed couplings on 
the branes. 
 
The dilaton in this region will be absent, or equivalently it 
will have zero coupling ${\cal C}_D=0$ to the branes because ${\cal K}_1 \to 
\infty$. 
 
The radion will have again positive kinetic term and its coupling will  
be bounded as following (see Fig. \ref{couplings}): 
\begin{equation} 
0 \leq g_R^{(r)} \lesssim e^{-2|k_1|r} \, , \qquad g_R^{(0)} \approx  1 
\end{equation} 
 
It is approximately constant on the central brane with strength equal 
to the one of the graviton  and is weakly coupled to the moving brane 
with coupling interpolating between zero in the 
inverted GRS case with $k_2 \to 0$ to a central value as the tension of 
the moving brane becomes infinitely large.

\section{Discussion and conclusions}   
In the previous sections we have considered the 
infinitesimal branes motion consistent 
with our linearized treatment of Einstein 
equations. The "small" motion takes place as 
a fluctuation around the unperturbed position 
$y=r$. It is interesting to ask what would happen in the 
generic situation when the small perturbation 
condition is relaxed. In this case the radion 
excitations will be defined as the 
perturbation around the a priori determined time 
varying positions of the branes in the bulk.

The motion of the brane will be an additional source of energy and 
momentum in the bulk and will excite all gravity excitations between 
which the radion as well. It would be interesting to find a actual 
solution of a moving brane in the bulk and study the dynamics of the 
radion field. In  \cite{Binetruy:2001tc} the analogous idea for the 
dilaton was studied. In the case of the radion there is not a simple 
solution preserving Poincar\'{e} or (A)dS invariance on the 
branes when they are moving. The dynamics of the radions in a system of 
the moving branes is more subtle than the ones of the dilaton and we
hope to address this problem in the future.
 
In conclusion, in this paper we presented the dilaton and radion 
dynamics in a flat brane system in warped bulk. We showed how one 
could calculate the effective action for these modes for a general three brane  
compact model. In the following we examined the kinetic term 
coefficients for these modes and for the sake of simplicity we sent 
the third brane at infinite distance. The physics was in accordance 
with our intuition that a positive tension brane has positive kinetic 
term and a negative tension one gives rise to a ghost radion. We 
calculated the  radion and studied it for  all possible cases of brane 
tension combinations. Additionally, we presented how the tensor 
gravity excitations behave in the above regions. Finally, we 
speculated about the motion of the brane being a source for the radion  
field.

\textbf{Acknowledgments:} We would like to thank  Thibault Damour,
Panagiota Kanti, Graham G. Ross and Arkady Vainshtein  for  very
stimulating discussions. S.M.'s work is supported by the Hellenic State   
Scholarship Foundation (IKY) \mbox{No.    
8117781027}. A.P.'s work is supported by the Hellenic State Scholarship   
Foundation (IKY) \mbox{No. 8017711802}. This work   is   
supported in part by the PPARC rolling grant PPA/G/O/1998/00567, by   
the EC TMR grants  HRRN-CT-2000-00148 and  HPRN-CT-2000-00152.

\def\theequation{A.\arabic{equation}}   
\setcounter{equation}{0}   
\vskip0.8cm   
\noindent   
\centerline{\Large \bf Appendix}   
\vskip0.4cm   
\noindent   
It is convenient to define a new variable $z$ defined by   
\begin{equation}   
\frac{1}{a(y)} = \frac{dz}{dy} \quad .   
\end{equation}   
In the coordinates $(x,z)$ the metric (\ref{pert}) is conformal to a flat    
perturbed metric $\bar{G}_{AB}=   
\eta_{AB} + H_{AB}$   
\be   
\begin{split}   
ds^2 &= a^2 \left[\bar{G}_{\mu \nu}  \, dx^\mu dx^\nu + \bar{G}_{zz} \, dz^2 \right] =    
a^2 \left[\eta_{MN} + H_{MN}  \, dx^M dx^N \right]  \\   
H_{\mu \nu} &= \varphi_1(x,z)  \,  \eta_{\mu \nu} + 2 \epsilon(z) \,    
\de_\mu \de_\nu f_2(x) +    
h_{\mu \nu}(x,z) \\   
H_{zz} &= \varphi_2(x,z)  \end{split}   
\label{pert1}   
\end{equation}   
Inserting (\ref{pert1}) in (\ref{act}) and taking into account the equation   
of motion satisfied by $a$ one gets   
\begin{equation}   
\begin{split}   
S_{eff}& = \int d^4x \; {\cal L}_{eff} =  \int d^4x  \; 2 M_5^3\int dz     
\left[a^3  {\cal L}_{PF}(h) + \frac{a^3}{4}    
\left[(\de_z h)^2 - \de_z h_{\mu \nu} \, \de_z h^{\mu \nu}\right]+ {\cal L}_{\varphi} +    
{\cal L}_{h \varphi}    
\right] \quad . \\   
 {\cal L}_{\varphi} &={\cal L}_1  + {\cal L}_2 +  {\cal L}_{12} \quad ;   
\end{split}   
\end{equation}   
Where   
\begin{equation}   
\begin{split}    
{\cal L}_{h \varphi} &= \left[a^3 \left(\varphi_1 + \frac{1}{2} \varphi_2    
\right)+ f_2 \frac{d}{dz}    
\left(\epsilon^\prime a^3 \right)    
\right] \partial_\mu \partial_\nu h^{\mu \nu} \\   
&- \left[a^3 \left(\varphi_1 +   
 \frac{1}{2}\varphi_2 \right) + f_2     
\frac{d}{dz} \left(\epsilon^\prime a^3 \right) \right] \Box h    
+\frac{3}{2} \frac{d}{dz} \left(a^2 a^\prime \varphi_2 - a^3    
\varphi_1^\prime \right) h \quad ;    
\end{split}   
\end{equation}   
\begin{equation}   
{\cal L}_1 = \de_\mu f_1 \de^\mu f_1 \, \frac{3}{2} a^3 \left(Q^2 + Q q \right)   
+ f_1^2 \left(3 a^3 {Q^\prime}^2 + 3 {a^\prime}^2 a q^2 - 6 a^2 a^\prime    
Q^\prime q \right) \; ; 
\end{equation}   
\begin{equation}   
\begin{split}   
{\cal L}_2 &= \de_\mu f_2 \de^\mu f_2 \left(\frac{3}{2} a^3 B^2 + 3 a^3 AB   
-3 a^3 B^\prime \epsilon^\prime + 6 a^2 a^\prime \epsilon^\prime A \right)\\   
&+ f_2^2 \left(3 a^3 { B^\prime}^2 + 12 {a^\prime}^2 a A^2 - 12 a^2 a^\prime   
A B^\prime \right) \; ;  
\end{split}   
\end{equation}   
\begin{equation}   
\begin{split}   
{\cal L}_{12} &= \de_\mu f_1 \de^\mu f_2 \left(\frac{3}{2} a^3 B q + 3a^3 A Q   
+3a^3B Q - 3 a^3 \epsilon^\prime Q^\prime + 3 a^2 a^\prime \epsilon^\prime q \right) \\   
&+ f_1 f_2 \left[6 a^3  B^\prime Q^\prime + 12 {a^\prime}^2 a A q - 6 a^2   
a^\prime \left(2 A Q^\prime + q  B^\prime \right) \right] \; ;   
\end{split}   
\end{equation}   
and ${\cal L}_{PF}(h)$ is the 4D Pauli-Fierz Lagrangian for $h$   
\begin{equation}   
{\cal L}_{PF}(h) = \frac{1}{2} \de_\nu h_{\mu \alpha} \, \de^\alpha    
h^{\mu \nu} -    
\frac{1}{4} \de_\mu h_{\alpha \beta} \, \de^\mu h^{\alpha \beta}    
- \frac{1}{2} \de_\alpha h \, \de_\beta h^{\alpha \beta} +\frac{1}{4}    
\de_\alpha h \, \de^\alpha h \, .   
\end{equation}   
Differentiation with respect of $z$ is denoted with a prime. The absence of mixing   
terms in ${\cal L}_{eff}$ yields the following constraints   
\begin{eqnarray}   
&&A(z) = \frac{a B^\prime}{2 a^\prime} \, , \qquad \frac{d}{dz}    
\left(B a^2 \right) + \frac{2 a^\prime}{a^2} \, \frac{d}{dz} \left(a^3    
\epsilon^\prime \right) = 0 \;  \label{c1}\\   
&& Q(z) = c \, a^{-2} \, , \qquad q(z) = -2 c \, a^{-2}, \quad c \text{ is a   
constant} \; ; \label{c2}\\   
&& \int dz \, a A(z) = 0 \quad . \label{c3}   
\end{eqnarray}   
Eqns. (\ref{c1})-(\ref{c3}) give   
\begin{equation}   
\begin{split}   
{\cal L}_{eff} &=  2 M_5^3 \int dz  \Big \{a^3  {\cal L}_{PF}(h) +    
\frac{a^3}{4} \left[(\de_z h)^2 - \de_zh_{\mu \nu} \, \de_zh^{\mu \nu}    
\right] \\   
&- \frac{3}{2}c^2  a^{-1} \de_\mu f_1 \de^\mu f_1 + \frac{3}{4}    
\frac{a^2}{a^\prime} \, \frac{d}{dz} \left(B^2 a^2 \right)  \, \de_\mu f_2   
\de^\mu f_2 \Big \} \quad .   
\end{split}   
\label{eff}   
\end{equation}   
In particular the effective Lagrangian ${\cal L}_{Scal}$ for the dilaton    
$f_1$ and the radion   
$f_2$  is   
\begin{equation}   
\begin{split}   
{\cal L}_{Scal} &= {\cal K}_1 \, f_1 \Box f_1 +  {\cal K}_2 \, f_2 \Box f_2 \\   
& {\cal K}_1 =  2 M^3_5 \, \frac{3}{2}c^2 \int_{- L}^{L} a^{-2}    
\, dy \\   
& {\cal K}_2 = - 2 M^3_5 \, \frac{3}{4} \int_{- L}^{L} a    
\left(\frac{da}{dy} \right)^{-1} \, \frac{d}{dy} \left( B^2 a^2 \right) \, dy   
\quad ,   
\end{split}   
\end{equation}   
with   
\begin{equation}   
\begin{split}   
&\frac{d}{dy} \left(B a^2 \right) + 2 a^{-1} \, \frac{da}{dy} \, \frac{d}{dy}    
\left(a^4 \de_y \epsilon \right) = 0 \quad ; \\   
& \int_{- L}^{L} dy \; a \left(\frac{da}{dy} \right)^{-1} \,    
\frac{dB}{dy} = 0 \quad .   
\end{split}   
\end{equation}   
As a result only $\de_y\epsilon(0)$, $\de_y\epsilon(r)$ and 
$\de_y\epsilon(L)$ enter the radion effective action.

\end{document}